\pdfoutput=1
\documentclass[a4paper,11pt]{article}
\usepackage{jcappub_mod}

\usepackage[english]{babel}
\usepackage{lmodern}
\usepackage[T1]{fontenc}
\usepackage{fontawesome}

\usepackage{microtype}
\usepackage{soul}
\usepackage{xspace}
\usepackage{textcomp}
\usepackage{booktabs}
\usepackage{cleveref}
\crefname{figure}{figure}{figures}
\usepackage{relsize}

\usepackage{amsmath}
\usepackage{cancel}
\usepackage{siunitx}
\DeclareSIUnit{\kpc}{kpc}
\usepackage[version=4]{mhchem}

\usepackage[dvipsnames]{xcolor}
\hypersetup{colorlinks=true, linktoc=all, linkcolor=Blue, citecolor=Red, urlcolor=Blue}

\usepackage{tikz}
\usetikzlibrary{patterns}
\usetikzlibrary{decorations.pathmorphing}
\usetikzlibrary{decorations.markings}
\usetikzlibrary{angles}
\usetikzlibrary{quotes}

\graphicspath{{figures/}}

\newcommand{\cpp}{C\nolinebreak\hspace{-.05em}\raisebox{.3ex}{\relsize{-2}+}\nolinebreak\hspace{-.05em}\raisebox{.3ex}{\relsize{-2}+}\xspace}
\newcommand{\code}[1]{\texttt{#1}\xspace}
\newcommand{\gamalps}{\code{gammaALPs}}

\newcommand{\mtx}[2]{#1_\text{#2}}

\newcommand{\nmm}[2]{\newcommand{#1}{\ensuremath{#2}\xspace}}

\nmm{\dSN}{d}
\nmm{\ddec}{d_a}
\nmm{\tdec}{t_a}
\nmm{\tem}{\mtx{t}{em}}
\nmm{\tgeo}{\mtx{t}{geo}}
\nmm{\dg}{d_\gamma}
\nmm{\tg}{t_\gamma}
\nmm{\Eg}{E_\gamma}
\nmm{\EgR}{E_{\gamma,0}}
\nmm{\denv}{\mtx{r}{env}}
\nmm{\Aeff}{\mtx{A}{eff,$j$}}
\newcommand{\grc}{{\color{lightgray}c}}
\newcommand{\lorentz}{\begin{pmatrix} \beta \gamma & \gamma & 0 & 0\\ \gamma & \beta \gamma & 0 & 0 \\ 0 & 0 & 1 & 0 \\ 0 & 0 & 0 & 1 \end{pmatrix}}

\nmm{\ma}{m_a}
\nmm{\gagg}{g_{a\gamma}}
\nmm{\pag}{\overline{P}_{a\gamma}}
\nmm{\ks}{\mtx{\kappa}{s}}

\DeclareMathAlphabet{\mathpzc}{OT1}{pzc}{m}{it}
\newcommand{\coords}[1]{\mathpzc{#1}}
\newcommand{\dd}{\mathrm{d}}

\newcommand{\ee}{\mathrm{e}}
\newcommand{\vc}[1]{\mathbf{#1}}
\newcommand{\fourvc}[3]{\begin{pmatrix} #1 \\ #2 \\ #3 \\ 0 \end{pmatrix}}
\nmm{\lG}{\mtx{\lambda}{G}}
\nmm{\lP}{\mtx{\lambda}{P}}

\newcommand{\obp}{O+\xspace}
\newcommand{\chp}{Ch+\xspace}

\tikzset{photon/.style={decorate, decoration={snake}, draw=red}}

\definecolor{Blue}{RGB}{0,0,122}
\definecolor{Red}{RGB}{173,42,26}
\definecolor{gold}{HTML}{FFD700}
\definecolor{steelblue}{HTML}{4682B4}
\newcommand{\updated}[1]{#1}

\def\apj{ApJ}%
\def\jcap{JCAP}%
\def\mnras{MNRAS}%
\def\prd{Phys.\ Rev.\ D}%
\def\prl{Phys.\ Rev.\ Lett.}%
\def\solphys{Sol.\ Phys.}%
\def\nat{Nature}%
\def\iaucirc{IAU~Circ.}%
\def\jgr{J.\ Geophys.\ Res.}%
\def\memsai{Mem.\ Soc.\ Astron.\ Italiana}%
\title{Updated constraints on axion-like particles from temporal information in supernova SN1987A gamma-ray data}
\author[\,a,*]{Sebastian Hoof\begin{NoHyper}\note[*]{Now at Dipartimento di Fisica e Astronomia ``Galileo Galilei,''
{Universit\`a} degli Studi di Padova, and Istituto Nazionale di Fisica Nucleare -- Sezione di Padova in Padua, Italy.}\end{NoHyper}}
\author[\,b]{\& Lena Schulz}

\affiliation[a]{Institut f\"ur Theoretische Teilchenphysik (TTP), Karlsruher Institut f\"ur Technologie (KIT),\\
76128~Karlsruhe, Germany}
\affiliation[b]{Institut f\"ur Astrophysik und Geophysik (IAG), Georg-August-Universit\"at G\"ottingen,\\ Friedrich-Hund-Platz~1,
37077~Göttingen, Germany}

\emailAdd{hoof@pd.infn.it}
\emailAdd{lena.schulz@stud.uni-goettingen.de}

\preprintnumber{TTP22-072}
\arxivnumber{2212.09764}

\abstract{We revise gamma-ray limits on axion-like particles~(ALPs) emitted from supernova SN1987A based on Solar Maximum Mission data. We improve and simplify the computation of the expected gamma-ray signal from ALP decays, while also extending it to non-instantaneous ALP emission. For the first time we make use of the temporal information in the data to update the associated ALP-photon coupling limits. For ALP decays, our updated likelihood only mildly affects the limit compared to previous works due to the absorption of gamma rays close to SN1987A. However, for ALP conversions in the Galactic magnetic field, temporal information improves the limit on the ALP-photon coupling by a factor of 1.4.~\href{https://github.com/sebhoof/snax}{\faGithub}}

\begin{document}
\maketitle
\flushbottom

\section{Introduction}\label{sec:intro}

Axion-like particles~(ALPs) can arise as (pseudo-)Nambu--Goldstone bosons, associated with the breaking of a global $\mathrm{U}(1)$ symmetry (see e.g.\ refs~\cite{1002.0329,2012.05029} for reviews).
They may also appear in string theory compactifications~\cite{10.1016/0370-2693(84)90422-2,hep-th/0605206}, residing in the so-called ``axiverse'' with masses potentially spanning many orders of magnitude~\cite{0905.4720,1004.5138,1206.0819}.
Recently calculated, explicit mass spectra in type~IIB string theory confirm this general picture~\cite{2011.08693,2103.06812}.
In particular, ALPs from string theory need not be as light as their namesake, the QCD axion~\cite{1977_pq_axion1,1977_pq_axion2,1978_weinberg_axion,1978_wilczek_axion}, but can be much heavier.
In this work, we are specifically interested in masses up to the GeV scale.

In any case, constraining ALPs across different mass scales is evidently challenging.
Matters are also complicated by the fact that ALPs -- unlike QCD axions -- need not solve the Strong~CP problem, thus lacking a well-defined connection to QCD.
However, \updated{thanks to the realignment mechanism~\cite{Preskill:1982cy,Abbott:1982af,Dine:1982ah,Turner:1983he,Turner:1985si}}, ALPs are still excellent dark matter candidates~\cite[e.g.][]{1201.5902} and may also couple to photons.
This enables a large ensemble of experimental searchers to look for them in the laboratory and using astrophysical and cosmological probes (see e.g.\ ref.~\cite{1801.08127} for a review of ALP searches).

Particularly important events in astrophysics are supernovae~(SNe), such as the core-collapse supernova SN1987A in the Large Magellanic Cloud.
As we revisit in this work, SN1987A is unique in allowing us to constrain ALPs emitted during the SN on very different mass scales:

On the one hand, it is well known that the extreme path length in astrophysical settings can lead to strong constraints from particle decay~\cite{10.1103/PhysRevLett.39.784}.
The possibility of ALPs decaying into photons has been used in the past to place limits on couplings of ``heavy'' ALPs in the keV--GeV range~\cite{10.1103/PhysRevLett.62.509,1009.5714,1702.02964,2109.03244,2205.07896} and, similarly, on neutrinos~\cite{10.1103/PhysRevLett.62.505,10.1103/PhysRevLett.62.509,10.1016/0927-6505(93)90004-W,astro-ph/9601104}.

On the other hand, ``light'' ALPs with masses $\ma \lesssim \SI{1}{\nano\eV}$ may be converted e.g.\ inside the Galactic magnetic field into photons, which could then be detected.
This has been used to place some of the most competitive limits on the ALP-photon coupling in this mass range~\cite{astro-ph/9605197,astro-ph/9606028,1410.3747}.

Apart from considering individual SNe, it has also been pointed out that ALP emission from all past SNe gives rise to a diffuse supernova axion background~\cite{1110.6397}, which can be searched for via the conversion or decay of ALP into gamma-ray photons~\cite[e.g.][]{2006.06722,2008.11741}.

In this work, we improve the ALP-photon limit by including the available temporal information contained in the Solar Maximum Mission~(SMM) gamma-ray data.
We first describe the construction of our updated likelihood in \cref{sec:derivation} and \cref{app:digitisation}.
In particular, the theoretical computation of the expected ALP-induced gamma-ray flux from SN1987A is revisited in \cref{sec:signal} and \cref{app:derivation}, where we make further progress in the analytical formalism and extend the previous results to non-instantaneous ALP emission.
We present limits derived from our updated likelihood in \cref{sec:results}, comparing them to previous works and discussing the differences.
Finally, we conclude with an outlook and additional comments in \cref{sec:conclusions}.

The digitised data sets and computational routines used in this work are available on Github at \url{https://github.com/sebhoof/snax}.

\section{Constructing the updated likelihood}\label{sec:derivation}

Construction of the likelihood function (presented in \cref{sec:likelihood}) requires us to understand the available data and instrument response~(\cref{sec:data}), to select a sensible background model~(\cref{sec:bkg}), and to compute the expected gamma-ray signals~(\cref{sec:signal}).

\subsection{SN1987A observations}\label{sec:data}

Supernova SN1987A was observed in the Large Magellanic Cloud at Galactic coordinates of $l = \SI{279.703}{\degree}$ and $b=\SI{-31.937}{\degree}$ in February 1987 by telescopes in different locations~\cite{1987IAUC.4316....1K}.
The star Sanduleak~-69~202 was identified as the supernova progenitor~\cite{1987IAUC.4317....1H,10.1038/328318a0} at an estimated distance of $\dSN = \SI{54.1 \pm 1.2}{\kpc}$~\cite{1998MmSAI..69..225P,astro-ph/0309416}.

\paragraph{Neutrino data.}
In addition to the telescope observations, three neutrino observatories saw a neutrino burst around the same time: Kamiokande~II~\cite{10.1103/PhysRevLett.58.1490,10.1103/PhysRevD.38.448}, Irvine-Michigan-Brookhaven (IMB)~\cite{10.1103/PhysRevLett.58.1494,10.1103/PhysRevD.37.3361}, and Baksan~\cite{1987PZETF..45..461A,10.1016/0370-2693(88)91651-6}.
The neutrino data has been analysed in various studies~\cite[e.g.][]{astro-ph/0107260,0810.0466}, which find that the measurements are consistent with the first neutrino measured in each detector arriving simultaneously (with an uncertainty of less than a second).
Since the IMB detector had by far the most accurate clock, the weighted average of arrival times of the first neutrino is essentially identical to the IMB value, which is $t_\nu = \SI{27341.37 \pm 0.05}{\s}$ after 00:00:00\,UTC on 23 February 1987~\cite{10.1103/PhysRevD.37.3361}.

\paragraph{SMM/GRS data.}
Around the time of the neutrino burst, the gamma-ray spectrometer (GRS)~\cite{10.1007/BF00151381} aboard the SMM satellite was operational.
As shown in ref.~\cite[Fig.~2]{10.1103/PhysRevLett.62.505}, and as described in ref.~\cite{10.1016/0927-6505(93)90004-W}, the GRS took around \SI{223}{\s} of data following the arrival of the first neutrino.
Afterwards the GRS went into calibration mode for about \SI{10}{\min} before taking data for another \SI{15}{\min} or so.
After this second data-taking interval, the detector was switched off while transiting through the South Atlantic radiation anomaly.

The authors of ref.~\cite{10.1016/0927-6505(93)90004-W} decide against using the data from the second data-taking interval due to concerns about the background model.
While we think that it would still have been interesting to analyse it, we were unfortunately unable to obtain additional data despite a number of enquiries.
Only GRS data associated with solar flares appears to have been designated for long-term storage.\footnote{The SMM data archive is available at \url{https://umbra.nascom.nasa.gov/smm/}.}
This is also unfortunate in light of the slight discrepancies between the digitised data sets (up to 5\% shifts), which we discuss together with our digitisation procedure in \cref{app:digitisation}.

\begin{figure}
    \centering
    \includegraphics[width=6in]{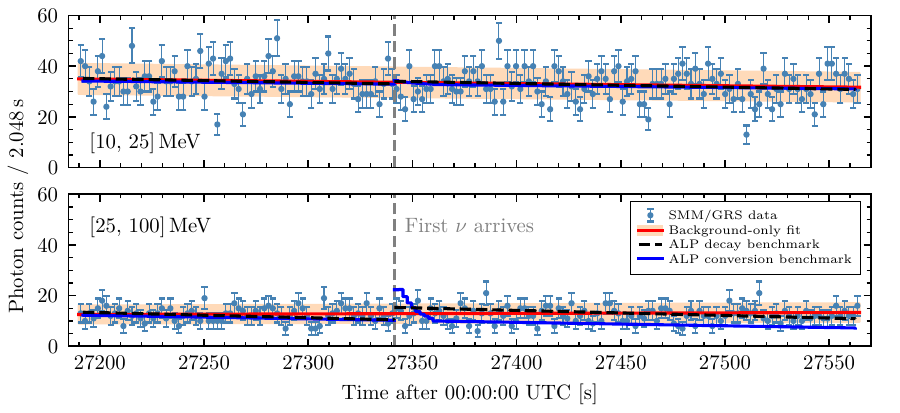}
    \caption{Photon counts data from SMM/GRS (blue dots and error bars) for two energy bands from ref.~\cite[Fig.~1]{10.1016/0927-6505(93)90004-W}. We also show the best-fitting background-only model (red lines and error bands), obtained from the data before the arrival of the first neutrino at $t_\nu$ (grey dashed vertical line). Two strongly disfavoured ($\lP \simeq \Delta\chi^2 = 25$) ALP benchmark models are shown for illustrative purposes: $\ma = \SI{1}{\MeV}$, $\gagg = \SI{2.26e-11}{\GeV^{-1}}$ (dashed black line) and $\ma = \SI{e-10}{\eV}$, $\gagg = \SI{6.55e-12}{\GeV^{-1}}$ (blue line).}
    \label{fig:data}
\end{figure}
In \cref{fig:data} we show the data for two of the available energy bands, which we digitised from the literature~(see \cref{app:digitisation} for details).
We do not include data from the \SIrange{4.1}{6.4}{\MeV} band since it has a negligible effect on our results due to its narrow range.
To a lesser extend this is also true for the \SIrange{10}{25}{\MeV} band, as illustrated by the benchmark models in \cref{fig:data}.

\paragraph{The GRS effective area.}
The GRS was facing the Sun during the neutrino burst, meaning that gamma rays from SN1987A had to penetrate the walls of the spacecraft in order to reach the detectors.
This reduced the effective detector area \Aeff for all energy bands~\cite{10.1016/0927-6505(93)90004-W}.
Still, observations of the \SI{847}{\keV} \ce{^56Co} line~\cite{10.1038/331416a0} demonstrate that the GRS was technically capable of detecting a gamma-ray burst at \si{\MeV} energies.

Estimating the effective area is nonetheless one of the major sources of uncertainties in limits derived from the GRS data set. 
Monte Carlo~(MC) simulations have been performed to estimate the effective area of the detector~\cite[Fig.~1(A)]{1985ICRC....5..474C}~(see also ref.~\cite[Fig.~1]{10.1016/0273-1177(86)90127-4}).
These MC simulations agree within 20--30\%~\cite{1985ICRC....5..474C} with the Earth's gamma-ray albedo flux measurements~\cite{10.1029/JA086iA03p01265}.
While this gives an estimate of the uncertainty of \Aeff under ``normal'' operating conditions, the difference in viewing angle during SN1987A may introduce additional uncertainties~(see also the discussion in ref.~\cite[§4.2.4]{1410.3747}).

The authors of ref.~\cite{10.1103/PhysRevLett.62.505} quote effective areas of \SI{115}{\cm^2} and \SI{63}{\cm^2} for the \SIrange{10}{25}{\MeV} and \SIrange{25}{100}{\MeV} energy bands, respectively.
While it is not clear in how far these values reflect the complications of the measurement, it appears that a reduction compared to the nominal effective area has been taken into account (cf.\ refs~\cite{1985ICRC....5..474C,10.1016/0273-1177(86)90127-4}).

In contrast the authors of ref.~\cite{10.1016/0927-6505(93)90004-W} quote a \emph{larger} effective area of \SI{90}{\cm^2} for the \SIrange{25}{100}{\MeV} band, without providing details about the computation.
The authors also argue that, due to the high gamma-ray energies, the effective areas should be close to their nominal values.
Following this logic, we would have also expected larger \Aeff values for the other energy bands.

It is unfortunately not possible anymore to validate the calculations of \Aeff due to the lack of information provided.
We use the values quoted in the earlier work, ref.~\cite{10.1103/PhysRevLett.62.505}.
This is the more conservative choice, also allowing us to directly compare our results with most of the later literature.

\subsection{Background model}\label{sec:bkg}
As discussed in \cref{sec:data}, we assume that the arrival time of the first neutrino~($t_\nu$) coincides with the travel time of light to SN1987A.
The data can then be divided into an ``off'' and ``on'' measurement, where the ``off'' data is used to fit the background model nuisance parameters.

To analyse the available photon counting data, we use a Poisson likelihood.
For the ``off'' data, we have \updated{(up to a constant)}
\begin{equation}
   \log \mtx{L}{off} \equiv \log \, p(n_{ij} \, | \, b_{ij},\, t_i < t_\nu) = \sum_{j=1}^{2} \sum_{i=1}^{i_\nu - 1} \left(n_{ij} \log(b_{ij}) - b_{ij} \right) \, , \label{eq:nuisance_loglike}
\end{equation}
where $b_{ij}$ and $n_{ij}$ are the background model prediction and number of photon counts and in the $i$th time and $j$th energy bin, respectively, while $i_\nu$ denotes the index of the time bin that contains $t_\nu$.

Different background models were analysed in ref.~\cite{10.1103/PhysRevLett.62.505} by considering photon count data on the day before and after SN1987A.
For a timescale of $\SI{36}{\min}$ before the satellite went into calibration mode, the authors conclude that a quadratic background model should be used to describe the data.
However, over shorter timescales -- such as the $\SI{6}{\min}$ interval in \cref{fig:data} -- we find that a linear model is sufficient to describe the data.
We parameterise our linear ansatz for fitting the ``off'' data as
\begin{equation}
    b_{ij} = a_{j}^{(0)} + a_{j}^{(1)} \, \frac{t_i - t_\nu}{\Delta t} \, , \label{eq:bkg_model}
\end{equation}
where $t_i$ is the time at the centre of the $i$th time bin and $\Delta t = \SI{2.048}{\s}$.
Since the coefficients $a_{j}^{(0)}$ and $a_{j}^{(1)}$ only depend on the data in the $j$th energy bin, we can optimise the partial likelihoods for the $j$th energy bin independently instead of \cref{eq:nuisance_loglike}.

\begin{table}
    \centering
    \caption{Overview of parameters for the SMM/GRS properties and data. For each energy bin, we quote the effective area \Aeff and the best-fitting parameters of the background-only model, derived from fitting the ``off'' data.}
    \label{tab:summary}
    \begin{tabular}{cccrr}
    \toprule
    Index $j$ & Energy band [\si{\MeV}] & \Aeff [\si{\cm^2}] & \multicolumn{1}{c}{$\hat{a}_{j}^{(0)}$} & \multicolumn{1}{c}{$\hat{a}_{j}^{(1)}$} \\
    \midrule
    1 & \numrange{10}{25} & $115$ & $33.6512$ & $-0.0178$ \\
    2 & \numrange{25}{100} & $63$ & $12.8749$ & $0.0043$ \\
    \bottomrule
    \end{tabular}
\end{table}

We quote our best-fitting parameters for the background in \cref{tab:summary}, along with the respective effective areas for each energy bin.
\Cref{fig:data} shows the prediction of this background model for both the ``off'' and ``on'' regions of the data (solid red lines).
As in all previous works, we too find excellent agreement of all data with the background-only hypothesis.

\subsection{Signal prediction}\label{sec:signal}

Let us now compute the expected number of gamma rays from ALPs emitted during SN1987A.
While both conversion (see \cref{sec:conv}) and decay (see \cref{sec:decay}) processes come from the same ALP spectrum, the relevant mass scales at allowed couplings are separated by some twelve orders of magnitude.
The limits can thus be derived independently with the same likelihood, simply replacing the expressions for the expected number of photons $s_{ij}$ in the $i$th time and $j$th energy bin.

\subsubsection{Emission spectrum}\label{sec:alp_spectrum}
Axion-like particles can be produced in SNe via their interactions with fermions, nucleons, photons, or pions~\cite{10.1016/0370-2693(87)91710-2,10.1103/PhysRevLett.60.1793,10.1103/PhysRevD.38.2338,1410.3747,1906.11844,2005.07141,2008.04918,2107.12393,2109.03244} (see also ref.~\cite{2205.07896} for a discussion of loop-induced couplings).
In this work, we focus exclusively on ALP-photon interactions, for which Primakoff production~\cite{10.1103/PhysRev.81.899} and photon-photon coalescence~\cite{hep-ph/0006327,2008.04918} are the most relevant processes.

\updated{In the following we only consider Primakoff production, so let us discuss how including the coalescence process would affect our results.
To this end we can rely on coalescence rates computed in an upcoming study~\cite{2023_calore_sn1987a_alps},\footnote{\updated{We thank Eike M\"uller for making these results available to us.}} based on previous SN simulations.
The additional contribution to the ALP flux from the coalescence processes has two effects:
one is that the coalescence process starts dominating for ALP masses $\ma \gtrsim \SI{50}{\MeV}$, and we will thus obtain stronger limits on \gagg for these masses.
The other effect is that the bounds extend to slightly higher masses, namely up to $\ma \sim \SI{250}{\MeV}$ compared to $\ma \sim \SI{200}{\MeV}$ for Primakoff production only.

For the Primakoff-induced ALP flux, we interpolate the normalisation constant $C_1$, average energy $E_\ast$, and exponent~$\alpha$, tabulated in ref.~\cite[Table~1]{1410.3747}, using cubic splines.}
The axion emission spectrum for axion energy $E_a$ and emission time \tem is, in the limit of $\ma \rightarrow 0$, given by the parametric form \cite[Eq.~(2.11)]{1410.3747}
\begin{equation}
    \frac{\dd^2 N_a}{\dd\tem \dd E_a} \approx C_1(\tem; \, \gagg) \, \left(\frac{E_a}{E_\ast(\tem)}\right)^{\alpha(\tem)} \ee^{- (\alpha(\tem) + 1) E_a/E_\ast(\tem)} \, . \label{eq:alp_emission_spectrum}
\end{equation}
We then also fit the instantaneous emission spectrum to the parametric form proposed in ref.~\cite[Eq.~(7)]{1702.02964}
\begin{align}
    \frac{\dd N_a}{ \dd E_a} &= \int_{\SI{0.005}{\s}}^{\SI{18}{\s}} \! \dd \tem \; \frac{\dd^2 N_a}{\dd\tem \dd E_a} \approx C_2 \, \frac{E_a^2}{\exp(E_a/\mtx{T}{eff}) - 1} \, \sigma_0(E_a;\, \gagg,\, \ks) \label{eq:int_alp_emission_spectrum} \\
    \text{with} \quad \sigma_0(E_a;\, \gagg,\, \ks) &= \frac{\alpha_\text{\tiny EM}\gagg^2}{8} \left[\left(1 + \left(\frac{\ks}{2E_a}\right)^2\right) \log\left(\left(\frac{2E_a}{\ks}\right)^2 + 1\right)- 1\right] \, . \label{eq:massless_primakoff}
\end{align}
For the reference value of $\gagg = \SI{e-10}{\GeV^{-1}}$, we find $\hat{C}_2 = \SI{2.03e77}{\MeV^{-1}}$, $\mtx{\hat{T}}{eff} = \SI{31.3}{\MeV}$, and $\mtx{\hat{\kappa}}{s} = \SI{17.3}{\MeV}$.
While \updated{the value for} $C_2$ is slightly lower compared to ref.~\cite{1702.02964} -- which is likely due to our different interpolation method for the coefficients in \cref{eq:alp_emission_spectrum} -- the total number of emitted ALPs from fully integrating the spectra only differs by 1--2\%.

The spectrum for massive ALPs is then approximately obtained by replacing the Primakoff cross section for massless ALPs in \cref{eq:int_alp_emission_spectrum}, $\sigma_0$, with the corresponding expression for massive particles, $\sigma(E_a;\, \ma, \, \gagg,\, \ks)$, as discussed in ref.~\cite[Eq.~(9)]{1702.02964}.

\subsubsection{ALP conversion signal}\label{sec:conv}

\updated{Apart from decays, the \gagg coupling also allows for the mixing of ALPs and photons in the presence of external magnetic fields.
Reference~\cite{10.1103/PhysRevD.37.1237} was the first to correctly describe the evolution of the ALP-photon system in general magnetic field configurations.

For this work, we consider the Galactic magnetic field~(GMF), for which the models of either Jansson \& Farrar~(J\&F)~\cite{1204.3662} or Pshirkov et al.~(P+)~\cite{1103.0814} are typically used.
We choose the J\&F model to compare our results with the literature.
Note that the \gagg limits from the J\&F model are weaker by a factor of 2--3 compared to P+~\cite{1410.3747}, which makes the J\&F model a conservative choice.

There exist several software codes for computing the overall conversion probability from ALPs emitted from SN1987A into photons detected by the SMM satellite.
Amongst them are the \texttt{ALPro}~\cite{10.5281/zenodo.6137185} or \gamalps~\cite{10.5281/zenodo.4973513} packages, from which we choose the \gamalps as it includes the J\&F model by default.

Since the exact GMF configuration at the time of SN1987A is unknown, the overall ALP-photon conversion probability can only be computed on average by simulating different field configurations.
In particular, the \gamalps code varies the GMF and splits up the line-of-sight between the SMM satellite and SN1987A into a sufficiently large number of ``cells'' of size~$L$~\cite{2108.02061}.
Inside these cells, the local (transverse) magnetic field $B$ is assumed to be constant.
Splitting up the path into many cells also naturally implements the matrix formalism described in ref.~\cite{10.1103/PhysRevD.37.1237}, according to which the ALP conversion probability -- to leading order in \gagg\xspace -- is given by}
\begin{equation}
    P_{a\gamma} = \left(\frac{\gagg B L}{\mtx{\Delta}{osc}}\right)^2 \, \sin^2 \left(\frac{\mtx{\Delta}{osc} L}{2}\right) \quad \text{with} \quad \mtx{\Delta}{osc}^2 = \left(\frac{\mtx{\omega}{pl}^2 - \ma^2}{2\Eg}\right)^2 + (\gagg B)^2 \, , \label{eq:p_agamma}
\end{equation}
where $\mtx{\omega}{pl}$ is the plasma frequency of the medium inside the cell.
If the typical size and strength of magnetic fields are known, MC simulations of the magnetic field can be used to obtain an average conversion rate \pag from ALPs into photons.

For the benchmark case of $\ma \lesssim \SI{e-11}{\eV}$ and $\gagg = \SI{e-10}{\GeV^{-1}}$, the authors of ref.~\cite[§3]{1410.3747} find a conversion rate of $\pag = 0.09$, which is about a factor of 1.6 larger than the result of \gamalps.
We find a similar discrepancy for the resulting fluence, suggesting that we can successfully replicate the remaining calculations.
The authors of ref.~\cite{1410.3747} state that they closely follow ref.~\cite{1207.0776} in their computations, but do not provide additional details beyond this.
The differences could be due to the further improvements added after the publication of ref.~\cite{1207.0776}, which eventually led to the release of the \gamalps code. The latter has also been cross-validated by an independent calculation for ref.~\cite{1712.01839}.\footnote{We thank Manuel Meyer for making us aware of this.}

Since low-mass ALPs are highly relativistic ($\Eg \simeq E_a$ and $t \simeq \tem$) and convert into one photon each, we have
\begin{align}
    s_{ij} &= \frac{\Aeff}{4\pi\dSN^2} \, \int_{E_j}^{E_j^\prime} \! \dd E_\gamma \, \int_{t_i}^{t_i^\prime} \! \dd t \; \frac{\dd^2 N_\gamma}{\dd t \, \dd E_a} \nonumber \\
    & = \frac{\Aeff}{4\pi\dSN^2} \, \int_{E_j}^{E_j^\prime} \! \dd E_a \, \int_{t_i}^{t_i^\prime} \! \dd \tem \; \pag(E_a;\, \ma,\, \gagg) \, \frac{\dd^2 N_a}{\dd \tem \, \dd E_a} \, ,
\end{align}
where \pag is computed by \gamalps, and the ALP emission spectrum is given by \cref{eq:alp_emission_spectrum}.

\updated{Finally, note that recently connections of \cref{eq:p_agamma} to Fourier analysis have been explored~\cite{1808.05916,2107.08040}, which is also used in the \texttt{ALPro} code.
Furthermore, the simple ``cell'' model may be replaced by Gaussian random fields or full magnetohydrodynamic simulations~\cite{1703.07314, 2208.04333}.}

\subsubsection{ALP decay signal}\label{sec:decay}

The expected signal from decaying ALPs~\cite{1009.5714,1702.02964,2109.03244,2205.07896,2205.13549} or sterile neutrinos~\cite{10.1016/0927-6505(93)90004-W,astro-ph/9601104,Raffelt:1996wa} has been calculated before, with various degrees of analytical and numerical methods such as MC simulations/integration or quadrature.
We include a number of improvements compared to previous works, as described in detail in \cref{app:derivation}.
Here we only quote the final result, according to which the expected number of photons $s_{ij}$ is given by
\begin{align}
    s_{ij} &= \frac{\Aeff}{4\pi\dSN^2} \int_{E_j}^{E_j^\prime} \! \dd \Eg \; \int_{\mtx{E}{min}}^{\mtx{E}{max}^{(i)}} \! \dd E_a \; \frac{2}{\beta E_a} \, \left[\exp\left(-\frac{\mtx{t}{min}^{(i)}}{\gamma \tau_{a\gamma,0}}\right) - \exp\left(-\frac{\mtx{t}{max}^{(i)}}{\gamma \tau_{a\gamma,0}}\right)\right] \, \frac{\dd N_a}{\dd E_a} \label{eq:master_integral} \\
    E_\text{min} &= \Eg + \frac{\ma^2}{4\Eg} \, , \; \mtx{E}{max}^{(i)} = \Eg + \frac{m^2}{4 \Eg} \left( \frac{\dSN^2}{(t_i + 2 \dSN) \, t_i} + 1 \right) \, , \\
    \mtx{t}{min}^{(i)} &= \max\left\{ \tdec(t_i) \, , \, \denv/\beta \right\} \, , \; \mtx{t}{max}^{(i)} = \min\left\{ \tdec(t_i^\prime) \, , \, \tdec(\tgeo) \right\} \, , \\
    \tdec(t) &= \frac{t + \dSN}{2 \, (1 - \Eg/E_a)} \, \left[1 \pm \sqrt{1 - \left[1 - \left(\frac{\dSN}{t + \dSN}\right)^2\right]\frac{4 \Eg (E_a - \Eg)}{\ma^2}}\right] \, , \\
    \tdec(\tgeo) &= \frac{d}{2} \, \sqrt{ \frac{E_a^2}{\left(E_a -\Eg\right) \left( E_a - \Eg - \frac{\ma^2}{4\Eg}\right)} } \, .
\end{align}

Note that we assume an instantaneous ALP emission for the decay limits, i.e.\ ignore the ALP emission time \tem considered in \cref{app:derivation}.
The non-instantaneous ALP emission becomes more relevant for decays close to SN1987A, which are however already strongly excluded by data or happen within \denv.
For ALPs that \updated{decay further away from SN1987A}, the expected signal in the first few minutes becomes relatively flat (see e.g.\ \cref{fig:data}), meaning that a non-instantaneous ALP emission only affects the signal in the first few out of the 109 bins in the ``on'' region of the data.

\subsection{Updated likelihood}\label{sec:likelihood}

In addition to the nuisance likelihood $\mtx{L}{off}$ introduced in \cref{sec:derivation}, we are now in a position to write the complete likelihood as 
\begin{equation}
    \log L(\ma,\, \gagg,\, a_{j}^{(0)}, a_{j}^{(1)}) \equiv \log \mtx{L}{off} + \sum_{j=1}^{2} \sum_{i=i_\nu}^{i_\nu + 108} \left( n_{ij} \log(b_{ij} + s_{ij}) - (b_{ij}+ s_{ij}) \right) \, , \label{eq:loglike}
\end{equation}
where $s_{ij}$ is either the signal prediction from ALP conversions, computed in \cref{sec:conv}, or decays, computed in \cref{sec:decay}.

Since we are only interested in limits in the $(\ma,\, \gagg)$ plane, we can ``profile out'' the nuisance parameters by considering the log-likelihood ratio test statistic $\lambda$.
For the Poissonian likelihood, this is
\begin{equation}
    \lP (\ma,\, \gagg) \equiv -2 \left[ \log L(\ma,\, \gagg,\, \hat{a}_{j}^{(0)}, \hat{a}_{j}^{(1)}) - \log  \hat{\hat{L}} \right] \, ,
\end{equation}
where $\hat{a}_{j}^{(0)}$, $\hat{a}_{j}^{(1)}$ are estimates to \emph{locally} maximise $L$, i.e.\ given fixed \ma and \gagg, while $\hat{\hat{L}}$ is an estimate for the \emph{global} maximum of $L$.

This in contrast with e.g.\ ref.~\cite{1702.02964}, where a Gaussian approximation to the likelihood was used, assuming that all observed events in the \SI{223}{\s} measurement window and \SIrange{25}{100}{\MeV} energy bin are background, which leads to
\begin{equation}
    \lG (\ma,\, \gagg) \equiv \frac{1}{\sigma_b^2} \left. \left(\sum_{i} s_{ij} \right)^2\right|_{j=2} \, ,
\end{equation}
where we use $\sigma_b^2 = 1393$ to match ref.~\cite{1702.02964}.

\section{Results and discussion}\label{sec:results}

\begin{figure}
    \centering
    {
    \includegraphics[width=3in]{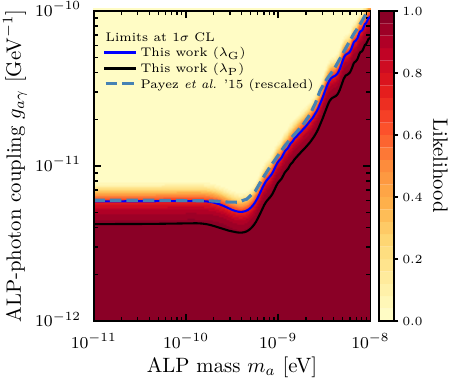}
    \hfill
    \includegraphics[width=3in]{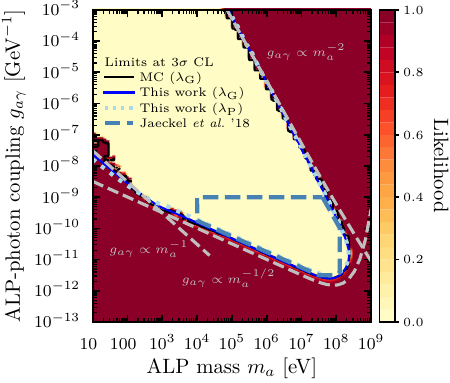}
    }
    \caption{Comparison of our results with the literature. \textit{Left:} Limits from ALP conversion in the Galactic magnetic field at $1\sigma$ confidence level~(CL). We show our results with \lG (blue line, density plot), the rescaled limit from ref.~\cite{1410.3747}~(dashed light blue line), and our results with \lP (black line). \textit{Right:} Limits from ALP decays at $3\sigma$ CL. We compare our results using \lG (blue line) and \lP (dotted light blue line) to those of ref.~\cite{1702.02964} (dashed light blue line) and the modified MC code from ref.~\cite{2205.13549} (black line, density plot). We also indicate the approximate scaling regimes of the limits~(dashed grey lines; see main text for details).}
    \label{fig:comparison}
\end{figure}

\Cref{fig:comparison} compares our results to the literature to validate and explain the differences in our updated limits, presented later in this section.

The left panel of \cref{fig:comparison} compares our ALP conversion limits~(black and blue lines) to ref.~\cite{1410.3747}, where we rescaled their limit~(dashed light blue line) to account for the difference in conversion rate $\pag$, as discussed in \cref{sec:conv}.
We set $\lambda \simeq \Delta\chi^2 = 1$ since ref.~\cite{1410.3747} considers a limiting photon fluence of \SI{0.6}{\cm^{-2}}.
While this corresponds to a $3\sigma$ upper limit for a \SI{10}{\s} window~\cite{10.1103/PhysRevLett.62.505}, the confidence level~(CL) is only about $1\sigma$ for the full \SI{223}{\s} window~(see also ref.~\cite{1702.02964}).

With the rescaling described before, we find excellent agreement with our computations and ref.~\cite{1410.3747} when using the simple Gaussian likelihood $\lG$ (blue line).
Once the timing information is included via $\lP$, the limit improves by a factor of about~1.4.
This is not unexpected since ALP emission from SN1987A mostly happens over a time window of \SI{20}{\s} or so.
Neglecting the time dependence essentially treats the signal as equally distributed across the whole time interval under consideration, while the actual standard deviation of the ALP emission time distribution is only about \SI{4}{\s}.
In other words: neglecting the temporal information can ``dilute'' the signal when distributed across many time bins.
This is also illustrated by the benchmark model (blue line) in \cref{fig:data}.

The right panel of \cref{fig:comparison} then shows our results for ALP decay limits (solid and dotted blue lines) compared to the previous results from ref.~\cite{1702.02964}~(dashed light blue lines).
For this purpose, we set $\lambda \simeq \Delta\chi^2 = 9$.
We also simplify the MC routines of ref.~\cite{2205.13549} along the lines of our derivations in \cref{app:derivation} and by introducing the popular MC integrator Python package \texttt{vegas}~\cite{2009.05112} as a more efficient integrator compared to brute-force MC simulations.
The results are shown as the density plot and black lines in the right panel of \cref{fig:comparison}.
Despite our improvements, we can see that MC simulations still struggle to correctly capture the low-mass region.
This is because, in the low-mass region, the acceptance fraction of the MC simulations are orders of magnitude smaller than for higher masses.
The authors of ref.~\cite{2205.13549} state that the number of direct MC simulations should be $\mathcal{O}(\num{e7})$.
However, they did not consider the low-mass region, where the number of MC simulations would have be increased according to the decrease in acceptance fraction. 

In any case, our updated likelihood \lP for ALP decays does not result in stronger limits despite containing additional temporal information.
This not due to e.g.\ the inclusion of the background nuisance parameters or other effects. In fact, the finer binning in time does not play much of a role since early ALP decay photons are reabsorbed in the envelope near SN1987A.
For later decays, the temporal distribution of the arrival photons becomes relatively flat during the first few minutes, thus not containing any useful timing information.
This is also illustrated by the benchmark model (dashed black line) in \cref{fig:data}.

For a better understanding of how the ALP decay limits in the left panel of \cref{fig:comparison} arise, we also indicate their approximate scaling behaviour (dashed grey lines and text) for \ma and \gagg, as previously discussed in ref.~\cite{1702.02964}.
In the regimes delimited by $d_a \sim \dSN$~($\gagg \propto \ma^{-1}$) and $t \sim \SI{223}{\s}$ ($\gagg \propto \ma^{-1/2}$) we set $E_a = \mtx{E}{avg}$, where the average ALP energy is $\mtx{E}{avg} \approx \SI{102}{\MeV}$ for ALPs with sub-MeV masses.
In the regime of too early decays, $d_a < \denv$~($\gagg \propto \ma^{-2}$), we set $E_a = E_{95}$, where $E_{95}$ is the 95th percentile of the ALP energy distribution.
For ALPs with sub-MeV masses, we find that $E_{95}= \SI{208}{\MeV}$.

\begin{figure}
    \centering
    {
    \includegraphics[width=3in]{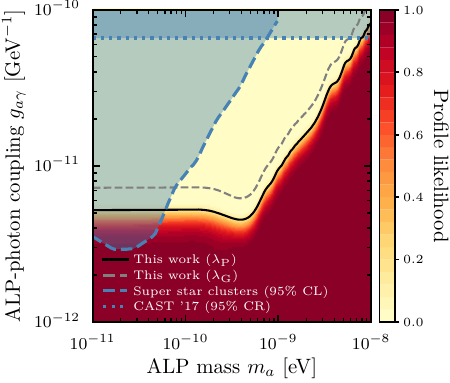}
    \hfill
    \includegraphics[width=3in]{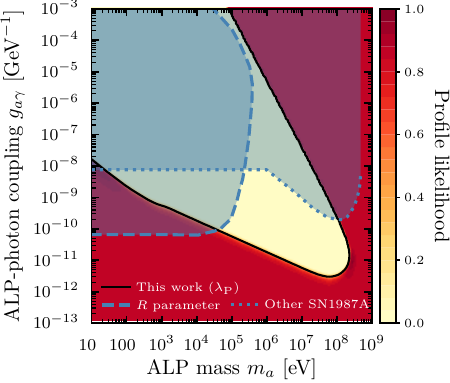}
    }
    \caption{Updated limits on the ALP-photon coupling for ALP-photon conversion (\textit{left}) and decays (\textit{right}) at the 95\% CL. For context we also show limits from the CAST helioscope~\cite{1705.02290} and conversion of ALPs produced in super star clusters in the Galactic magnetic field~\cite{2008.03305} (\textit{left}), and stellar evolution~($R$ parameter)~\cite{1406.6053} and other SN1987A processes (energy loss and duration of the neutrino burst)~\cite{2201.09890,2109.03244} (\textit{right}). We used the \texttt{AxionLimits} repository~\cite{zenodo_axionlimits} for some of the tabulated limits.}
    \label{fig:results}
\end{figure}
Finally, in \cref{fig:results} we show our updated limits using \lP (black lines) for the more standard 95\% CL to allow a direct comparison with other limits in the literature.
For ALP conversions (left panel), we also include the simpler \lG treatment to highlight that also the limits at this CL are still a factor of 1.4 stronger when using \lP.

\updated{Apart from complementing other limits, also note that ALP conversion after SN1987A excludes part of the parameter space where ALPs can explain the transparency of the Universe to gamma rays with energies $\gtrsim \SI{0.1}{\TeV}$~\cite{1302.1208,1704.05189} (not shown in \cref{fig:results}).}

To derive the 95\% CL threshold, we assume that \lP follows a $\frac12\chi_1^2 + \frac12\chi_2^2$ distribution, where $\chi_n^2$ is a $\chi^2$ distribution with $n$ degrees of freedom.
This generalisation~\cite{10.2307/2289471} of Chernoff's theorem~\cite{10.1214/aoms/1177728725} corresponds to $\lP \approx 5.14$, which is between the two-sided $\Delta\chi^2$ values for one ($\lP \approx 3.84$) and two ($\lP \approx 5.99$) degrees of freedom from Wilks' theorem~\cite{10.1214/aoms/1177732360}.
This takes into account that $\gagg = 0$ can lie on the boundary of the parameter space, which would make the ALP mass, $\ma > 0$, an unidentifiable parameter.
Despite this correction, parameter degeneracies and the ``look-elsewhere effect'' may further affect the \lP distribution.
Explicit MC simulations for a combination of ALP likelihoods have been performed to investigate these effects~\cite[appendix~D]{2007.05517}.
The results suggest that such generalisations of Wilks' theorem may offer a decent approximate interpolation, which do not require costly MC simulations.
In any case, the important point is that the inclusion of temporal information will lead to the same relative improvement of the limit regardless of the true $\lP$ threshold for a 95\% CL.

\section{Summary and outlook}\label{sec:conclusions}

We revise the limits on the ALP-photon coupling from ALPs emitted during supernova SN1987A, finding that temporal information can improve the limit from ALP conversions by a factor of~1.4.
A similar improvement for ALP decays is not possible since the photons from early ALP decays are absorbed in the envelope of SN1987A.
\updated{For later decays, the timing information is irrelevant as the photon signal becomes stretched out, meaning that it is essentially flat during the first few minutes after the arrival of the first neutrino.}

Still, we generalise the signal prediction from ALP decays to arbitrary emission and decay times, while making both analytical and numerical progress in evaluating the associated integrals.
Regardless of the strength of the limits, our updated likelihood approach is more realistic and complete than previous approaches, thanks to the inclusion of a background model, one additional energy bin and temporal information in a Poisson likelihood.

The digitised Solar Maximum Mission data for our updated likelihood and the Python/\cpp code for computing the signal prediction are publicly available on Github at \url{https://github.com/sebhoof/snax}.

Despite the improved limits from the statistical analysis in this work, one should keep in mind that there are sizeable uncertainties coming from the supernova emission model, where the predicted signals could change by an order of magnitude.
We also neglected the uncertainties from the distance to SN1987A (2.3\%), a possible systematic shift of the photon counts depending on the choice of digitised data set (5\%) and, most importantly, the effective area of the detector (estimated to be at least $\numrange{20}{30}\%$).
These effects could result in a relative uncertainty of $\sim \numrange{5}{7}\%$ on the location of the \gagg limit.

With this said, future work lies in improving the ALP emission predictions from supernova modelling.
Future nearby supernovae, such as the one predicted for the red supergiant Betelgeuse, will provide significantly more data and a drastically improved sensitivity to the ALP-photon coupling~\cite[e.g.][]{1609.02350,1702.02964}.
Given that even the limited amount of SN1987A data leads to some of the strongest constraints on ALPs to date, this presents an exciting prospect.
In such an event, the more detailed computation, analysis framework, and software code presented here will hopefully prove useful for probing ALP couplings across many orders of magnitude in ALP mass.

\acknowledgments

We thank Joerg Jaeckel, Manuel Meyer, and Edoardo Vitagliano for helpful discussions, Werner Collmar and Gerald H.\ Share for their efforts in trying to recover additional SMM/GRS data, \updated{and Eike M\"uller for helpful discussions and sharing ALP production rates from photon coalescence for comparison}.
We also acknowledge contributions by Csaba Bal\'azs and Marie Lecroq during an earlier, related project (see ref.~\cite{2205.13549}), \updated{the anonymous referee for helpful comments to improve the clarity of our manuscript}, and Tom\'as Gonzalo for suggesting the name for our software code.
SH was supported by the \textit{Deutsche Forschungsgemeinschaft} (DFG) through the Collaborative Research Center TRR~257 under Grant 396021762.
Parts of this work are based on LS's BSc thesis.
\updated{We made use of the \code{BibCom} tool~\cite{bibcom}.}

\appendix

\section{Digitisation of SMM/GRS data}\label{app:digitisation}

To extract the time-binned SMM/GRS data, we digitise Fig.~4 from ref.~\cite{10.1103/PhysRevLett.62.505} (``\chp'') and Fig.~1 from ref.~\cite{10.1016/0927-6505(93)90004-W} (``\obp'')  using the \texttt{Webplotdigitizer} tool~\cite{Rohatgi2022}.\footnote{Note that the quality of the journal's online version of ref.~\cite[Fig.~1]{10.1016/0927-6505(93)90004-W} is not suitable for digitisation, which is why we use a high-resolution scan of a physical copy of the article instead.}
The two data sets use a time binning of $\Delta t_{10} = \SI{10.24}{\s}$ and $\Delta t = \SI{2.048}{\s}$, respectively, while ref.~\cite[Fig.~4]{10.1103/PhysRevLett.62.505} gives us access to about \SI{3}{\min} of additional ``off'' data for all energy bins.
We confirm that our digitised data agrees with the binning stated in the papers since we find $\Delta t_{10}/5 = \SI{2.04 \pm 0.15}{\s}$ and $\Delta t = \SI{2.04 \pm 0.16}{\s}$.

Before explaining how we obtain our consensus data of integer photon counts, we note that we found a rather large discrepancy in the \SIrange{25}{100}{\MeV} energy bin, which cannot be explained by inaccuracies in the digitisation procedure.
In the time range where we can compare them, the number of photons in \obp is about 40\% larger than what we see in \chp.
It seems plausible to us that the authors of \obp did not have access to the actual photon count data but rather fluence data, to which they applied their higher value of the effective area of $\mtx{A}{eff,2} = \SI{90}{\cm^2}$~(cf.\ \cref{sec:data}).
To rectify this, we multiply the data in the \SIrange{25}{100}{\MeV} energy range of \obp with a factor of $63/90 = 0.7$ before proceeding.

We can then use the following estimators for the \obp photon counts in each bin: (i) the digitised data point, (ii) the average of the upper and lower error bar, and (iii) the square of half the length of the error bar.
Estimators (ii) and (iii) can be used since Fig.~1 of ref.~\cite{10.1016/0927-6505(93)90004-W} shows symmetrical error bars, suggesting that the authors use the Wald estimate for the uncertainty on $n$ measured photons, i.e.\ $n \pm \sqrt{n}$ for the $1\sigma$ interval.
Indeed, at least two of these estimators give the same rounded integer value for all data points.

Estimating the photon counts in each bin from \chp works in a similar way, except that the authors use asymmetric $2\sigma$ error bars, suggesting that they use a more rigorous approximation for their confidence intervals.
We find that the commonly used approximation $n + 2 \pm 2\sqrt{n + 1}$~\cite{10.1080/03610929408831336} agrees very well with the data.
We thus use the following estimators: (i) the digitised data point, (ii) a fit to the upper, and (iii) to the lower error bar, using the approximation above.
Again, at least two of these estimators give the same rounded integer value for all data points.

\begin{table}
    \centering
    \caption{Comparison of the digitised data sets from ref.~\cite[Fig.~4]{10.1103/PhysRevLett.62.505} (``\chp'') and ref.~\cite[Fig.~1]{10.1016/0927-6505(93)90004-W} (``\obp''). We quote the photon counts for our digitised data sets, the relative deviation to the \chp, and a cross check number of photons derived from information provided in Table~1 in ref.~\cite{10.1016/0927-6505(93)90004-W}.}
    \label{tab:validation}
    \begin{tabular}{lcrrrr}
    \toprule
    Data & \multicolumn{1}{l}{Energy band [\si{\MeV}]} & \multicolumn{1}{l}{\chp} & \multicolumn{1}{l}{\obp} & \multicolumn{1}{l}{Deviation} & \multicolumn{1}{l}{Ref.~\cite[Table~1]{10.1016/0927-6505(93)90004-W}} \\
    \midrule
    ``off'' & \numrange{10}{25} & 2434 & 2397 & $-1.5\%$ & \num{2434 \pm 1} \\
    & \numrange{25}{100} & 941 & 894 & $-5.0\%$ & \num{896 \pm 1} \\
    \midrule
    ``on'' & \numrange{10}{25} & 3673 & 3517 & $-4.2\%$ & \num{3590 \pm 2} \\
    & \numrange{25}{100} & 1421 & 1349 & $-5.1\%$ & \num{1366 \pm 2} \\
    \bottomrule
    \end{tabular}
\end{table}
How much are the digitised data sets from \chp and \obp in agreement?
To answer this question, we compare the ``on''~(\SI{223.232}{\s}) and the overlapping parts of the ``off'' data sets~(\SI{143.36}{\s}) in \cref{tab:validation}.
We find that the \obp data (after correction for the effective area) is systematically lower by up to about 5\%.
This might indicate e.g.\ a difference in the plotting routines used in the publications.
We checked that the difference is not due to a wrong calibration in our digitisation routines. 
As a result, trying to use data from both data sets -- to make use of the longer \chp ``off'' data time window and at the same time the more finely binned \obp  ``on'' data -- is not conservative.
The $\sim 5\%$ elevated background levels in \chp would lead to stronger bound from the \obp ``on'' data; although we find that the \gagg limit from ALP conversion would only be a factor of 1.5 stronger compared to not including temporal information (factor of 1.4 when only using \obp data).
We thus need to pick one of the two data sets and, since there is no definitive answer as to which digitised data set is more accurate, we decide to use the \obp data.

\section{Calculating the photon flux from astrophysical ALP decays}\label{app:derivation}

Here we provide a full derivation of the integral in \cref{eq:master_integral}, making use of the advantages of previous computations~\cite{10.1016/0927-6505(93)90004-W,astro-ph/9601104,Raffelt:1996wa,1009.5714,1702.02964,2109.03244,2205.07896,2205.13549} while using expressions valid for arbitrary decay times \tdec and a discussion of non-instantaneous ALP emission.

For simplicity, we set $c = \hbar = \mtx{k}{B} = 1$, except when emphasising the difference between times and lengths by reinstating ``$\grc$'' as a factor.

\subsection{Geometry and Lorentz boosts}

\begin{figure}
    \centering
    \begin{tikzpicture}[scale=1,rotate=0]
      \input{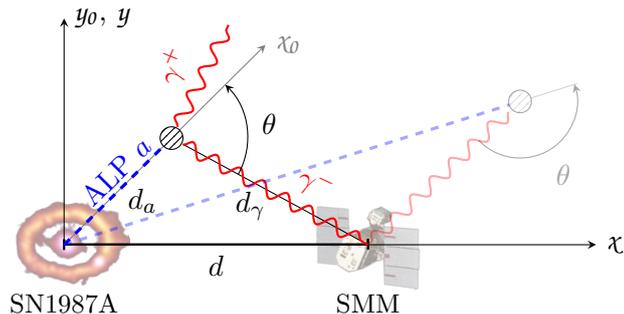}
    \end{tikzpicture}
    \caption{Geometry of ALP~(dashed blue line) decays into photons~(red lines) after SN1987A. \updated{The solid lines are labelled with the variables discussed in the main text, while the slightly transparent lines represent an ``extreme,'' improbable geometry}. Credit for the modified SSM satellite picture: G.\ Nelson/NASA (JSC image library; public domain) and modified SN1987A image: NASA/ESA, P.\ Challis, R.\ Kirshner, and B.\ Sugerman (Hubble image library; CC~BY~4.0).}
    \label{fig:geometry}
\end{figure}
\Cref{fig:geometry} shows the basic geometry of ALP decays after supernova SN1987A.
Without loss of generality, we may choose all ALP and photon paths to cross the $\coords{x}$-$\coords{y}$-plane.
In the ALP rest frame, the two decay photons are emitted back to back with energies of $\ma/2$ each.
The photon 4-momenta $p_{\gamma,0}^\pm$ in the ALP rest frame transform to the lab frame, where the ALP is moving with speed~$\beta$ in $\coords{x}$-direction, via the Lorentz boost $\Lambda$:
\begin{align}
    \Lambda = \lorentz \, , \quad p_{\gamma,0}^\pm = \EgR \fourvc{1}{\pm \cos(\theta_0)}{\pm \sin(\theta_0)} \mapsto p_\gamma^\pm = \EgR \fourvc{\beta\gamma \pm \gamma \cos(\theta_0)}{\gamma \pm \beta\gamma \cos(\theta_0)}{\pm \sin(\theta_0)}\, , \label{eq:boosts}
\end{align}
where
\begin{equation}
    \beta = \sqrt{1 - \left(\frac{\ma}{E_a}\right)^2} \, , \quad \gamma = \frac{E_a}{\ma} = \frac{1}{\sqrt{1 - \beta^2}} \, . \label{eq:beta_gamma}
\end{equation}

Since angles are defined via the 3-vector product, we find the emission angle $x^\pm \equiv \cos(\theta^\pm)$ from the $\coords{x}$-component of $\vc{p}_\gamma^\pm$. The photon energy $\Eg$ is, in turn, given by the 0-component of $p_\gamma^\pm$:
\begin{equation}
     x^\pm = \frac{\beta \pm x_0}{1 \pm \beta \, x_0} \, , \quad E_\gamma^\pm = \gamma (1 \pm \beta x_0) \, \EgR = \frac{1 \pm \beta x_0}{2} \, E_a \, , \label{eq:boosted_variables}
\end{equation}
where we used that $\EgR = \ma/2$ and defined $x_0^\pm \equiv \cos (\theta_0)$.

\updated{One consequence of the relativistic transformations was already pointed out in ref.~\cite{1702.02964}, namely that the decay photons for highly relativistic ALPs are emitted in a narrow forward cone in the lab frame.
This essentially means that e.g.\ geometries with ALP decays behind \dSN, such as the one shown as a slightly transparent path in \cref{fig:geometry}, may be neglected.
While geometries with ALP decays behind \dSN are possible for non-relativistic ALPs, they do not contribute much to the signal.
This is due to their low speed $\beta \ll 1$ compared to the short time window that we consider.}

\updated{Finally note that, due to the relabelling symmetry of the two photons, we may pick either sign \cref{eq:boosted_variables} as long as we include an overall multiplicity factor of two in what follows.
We choose to only discuss the ``$+$'' sign in \cref{eq:boosted_variables} to simplify the following derivations. While this choice does not correspond to the $\gamma^-$ photon shown in \cref{fig:geometry}, it makes it easier to compare to previous results in the literature, e.g.\ ref.~\cite[§12.4]{Raffelt:1996wa}.}

\subsection{Instantaneous ALP emission}\label{app:inst_alp_em}

We wish to obtain the signal prediction $s_{ij}$ in terms of photon counts by integrating the incoming photon flux over the $i$th time bin and $j$th energy bin for the effective detector area \Aeff.
We also need to consider the spectral distribution of axion energies $E_a$ and photon emission angles $x_0$ (whose distribution is known in the axion rest frame).
Similar to previous works, we also assume that photons from axions decaying within the envelope of the SN are fully absorbed~(cf.\ ref.~\cite{1702.02964}).
This leads us to
\begin{equation}
    s_{ij} = \int_{E_j}^{E_j^\prime} \! \dd \Eg \; \int_{t_i}^{t_i^\prime} \! \dd t \; \int_{\Aeff}^{} \! \dd A \; \frac{\dd^3 N_\gamma}{\dd \Eg \, \dd t \, \dd A} \; \Theta(\beta\tdec - \denv) \, . \label{eq:sij}
\end{equation}

Apart from these experimental parameters, we need to integrate over all (unobserved) variables, viz.\ the axion emission energies $E_a$, decay angles $x_0$ in the rest frame, and decay times $\tdec$.
Since $t$ will be related to the decay time $\tdec$, it is necessary to find an expression for $\tdec(t)$.
By applying the law of cosines to \cref{fig:geometry}, and using that $\cos(\pi - \alpha) = -\cos(\alpha)$, it follows for the \emph{path lengths} involved that
\begin{equation}
    \dSN^2 = \ddec^2 + \dg^2 - 2 \ddec \, \dg \, \cos (\pi - \theta) = (\beta \grc \, \tdec)^2 + (\grc \, \tg)^2 + 2 \beta \grc \, \tdec \, \tg \, x \, , \label{eq:decay_geometry}
\end{equation}
where we defined $x \equiv \cos(\theta)$ for convenience and all quantities are measured in the lab frame, i.e.\ the reference frame of the observing spacecraft.
We then define the measurement time $t$ in terms of other travel \emph{times}
\begin{equation}
    t \equiv \tdec + \tg - \dSN {\color{lightgray}/\grc} \, , \label{eq:time_variable}
\end{equation}
such that $t = 0$ coincides with the time measured after the arrival of the first (massless) neutrino, as discussed in the main text.

\updated{Replacing the photon path $\grc\,\tg$ in \cref{eq:decay_geometry} using \cref{eq:time_variable}, we obtain a quadratic polynomial in \tdec.
Further rewriting the polynomial with the help of \cref{eq:boosted_variables} and $\EgR = \ma/2$, we find that its two solutions are}
\begin{align}
    \tdec^{\pm} = \tdec^{\pm}(t) = \frac{t + \dSN}{2 \, (1 - \Eg/E_a)} \, \left[1 \pm \sqrt{1 - \left[1 - \left(\frac{\dSN}{t + \dSN}\right)^2\right]\frac{4 \Eg (E_a - \Eg)}{\ma^2}}\right] \, , \label{eq:tdec}
\end{align}
provided that the determinant is non-negative, which can be interpreted as a condition on $t$:
\begin{align}
    t/\dSN \leq \tgeo/\dSN \equiv \sqrt{\frac{E_a - \Eg}{E_a - \Eg - \frac{m^2}{4\Eg}}} - 1 \, .
\end{align}

\updated{To choose the physical solution for \tdec in \cref{eq:tdec}, we remind the reader that $\ma > 0$ is required for ALPs to decay into two photons.
As a consequence, $t = 0$ is only possible if $\tdec = 0$. Any decays with $\tdec > 0$ would lead to $t > 0$ due to the ALPs' subluminal speed $\beta < 1$.
Since $\tdec^{+}(0) \neq 0$ while $\tdec^{-}(0) = 0$, $\tdec(t) \equiv \tdec^{-}(t)$ is the physical solution.\footnote{\updated{The other, unphysical solution $\tdec^{+}$ has only a geometric interpretation.
The corresponding triangle can be obtained by mirroring the triangle in \cref{fig:geometry} at an axis perpendicular to the $\coords{x}$-axis at $\coords{x} = d/2$.}}

We also note that, in parts of the literature, the linear expansion of $\tdec^{-}$ has been used, which is~\cite[§12.4.4]{Raffelt:1996wa}
\begin{equation}
    \tdec^{-}(t) = \frac{2 E_a \Eg}{\ma^2} \, t + \mathcal{O}(t^2) \, . \label{eq:td_approx}
\end{equation}
However, we will see that the approximation in \cref{eq:td_approx} is not necessary and, in fact, late decays are relevant for parts of the parameter space.}

Knowing an expression for $\tdec(t)$ then allows a change of variables $t \mapsto \tdec$ in \cref{eq:sij}.
Together with the other unobserved variables, the relevant part becomes
\begin{equation}
    \int_{t_i}^{t_i^\prime} \! \dd t \; \frac{\dd^3 N_\gamma}{\dd \Eg \, \dd t \, \dd A} = \int_{\ma}^{\infty} \! \dd E_a \; \int_{-1}^{1}\!\dd x_0 \; \int_{\tdec(t_i)}^{\tdec(t_i^\prime)} \! \dd \tdec \; \frac{\dd^5 N_\gamma}{\dd x_0 \, \dd \Eg \, \dd \tdec \, \dd E_a \, \dd A} \, . \label{eq:integrand_1}
\end{equation}
Further expanding the integrand of \cref{eq:integrand_1} using the chain rule yields:
\begin{align}
    &\frac{\dd^2 N_\gamma}{\dd N_a \dd \EgR} \, \frac{\dd \EgR}{\dd \Eg} \, \frac{\dd^4 N_a}{\dd x_0 \, \dd \tdec \, \dd E_a \, \dd A} \\
    &= 2 \times \delta(\EgR - \ma/2) \times \frac{\dd \EgR}{\dd \Eg} \times \frac12 \times \frac{\ee^{-\tdec/\gamma \tau_{\text{tot},0}}}{\gamma \tau_{a\gamma,0}} \times \frac{\dd N_a}{\dd E_a} \times \frac{1}{4\pi\dSN^2} \label{eq:integrand_2}
\end{align}
where we used that -- in our case -- the total ALP lifetime equals the lifetime from photon decays, i.e.\ $\tau_{\text{tot},0} = \tau_{a\gamma,0} = 1/\Gamma_{a\gamma,0}$.\footnote{Note that this may not be true when other ALP interactions are present, such as an ALP-electron coupling~\cite[e.g.][]{2205.07896}, as already emphasised in ref.~\cite[§12.4.4]{Raffelt:1996wa} in the context of neutrino decays.}

Consider now the variable transform $x_0 \mapsto \EgR$.
By using \cref{eq:boosted_variables}, the resulting factor in the integrand combines with the remaining $\dd \EgR/\dd \Eg$ in \cref{eq:integrand_2} to an overall factor of
\begin{equation}
    \frac{\dd x_0}{\dd \EgR} \frac{\dd \EgR}{\dd \Eg} =  \frac{\dd x_0}{\dd \Eg} = \frac{2}{\beta E_a} \, .
\end{equation}
The transformation of the $x_0$~integral boundaries can be understood by writing them as $\Theta(x_0 + 1) \, \Theta(1 - x_0) = \Theta(1 - x^2_0)$.
Using \cref{eq:boosted_variables} and $\EgR = \ma/2$, one finds that
\begin{equation}
    \Theta(1 - x^2_0) = \Theta \bigg( 1 - \frac{1}{\beta^2} \, \Big( \frac{\Eg}{\gamma\EgR} - 1 \Big)^2 \bigg) = \updated{\Theta \Big( E_a - \Eg - \frac{\ma^2}{4\Eg} \Big)} \, . \label{eq:ea_min}
\end{equation}
\Cref{eq:ea_min} can be interpreted as a lower limit of the $E_a$ integral since $E_a \geq \Eg + \ma^2/4\Eg$~(cf.\ ref.~\cite[§12.4.5]{Raffelt:1996wa}).
This replaces the previous lower limit $E_a \geq \ma$ since $\Eg + \ma^2/4\Eg$ has a global minimum at $\Eg = \ma/2$ with value \ma.

Since \Aeff is an effective constant for the $j$th energy bin, we can put all ingredients \updated{together to find that}
\begin{align}
    s_{ij} &= \frac{\Aeff}{4\pi\dSN^2} \int_{E_j}^{E_j^\prime} \! \dd \Eg \; \int_{\mtx{E}{min}}^{\mtx{E}{max}^{(i)}} \! \dd E_a \; \frac{2}{\beta E_a} \, \frac{\dd N_a}{\dd E_a} \, \int_{\mtx{t}{min}^{(i)}}^{\mtx{t}{max}^{(i)}} \! \dd \tdec \; \frac{\ee^{-\tdec/\gamma \tau_{a\gamma,0}}}{\gamma \tau_{a\gamma,0}} \\
    &= \frac{\Aeff}{4\pi\dSN^2} \int_{E_j}^{E_j^\prime} \! \dd \Eg \; \int_{\mtx{E}{min}}^{\mtx{E}{max}^{(i)}} \! \dd E_a \; \frac{2}{\beta E_a} \, \left[\exp\left(-\frac{\mtx{t}{min}^{(i)}}{\gamma \tau_{a\gamma,0}}\right) - \exp\left(-\frac{\mtx{t}{max}^{(i)}}{\gamma \tau_{a\gamma,0}}\right)\right] \, \frac{\dd N_a}{\dd E_a}  \, , \label{eq:sij_simplified}
\end{align}
as long as $\mtx{t}{min}^{(i)} < \mtx{t}{max}^{(i)}$, where we defined
\begin{align}
    \mtx{E}{min} &= \Eg + \frac{\ma^2}{4\Eg} \, , \; \mtx{t}{min}^{(i)} = \max\left\{ \tdec(t_i) \, , \, \denv/\beta \right\} \, , \; \text{and} \; \mtx{t}{max}^{(i)} = \min\left\{ \tdec(t_i^\prime) \, , \, \tdec(\tgeo) \right\} \, .
\end{align}
We note that $\tdec(t_i) < \tdec(t_i^\prime)$ due to $t_i < t_i^\prime$,\footnote{\updated{Observe that $\tdec = \tdec^{-}$ in \cref{eq:tdec} is a product of two terms containing $t$, $t + d$ and the term in square brackets. Using $d > 0$, $E_a > \Eg$, and \cref{eq:ea_min}, it follows that both these terms are monotonic in $t$, meaning that \tdec is monotonic in $t$.}} while $\denv/\beta < \tdec(\tgeo)$ as long as $\denv < \dSN/2$.
We can also derive conditions on $E_a$ by comparing the other two remaining combinations of possible \tdec limits, which also improves the numerical convergence of the integral.
In practice, the easier condition on $E_a$ comes from $t_i < \tgeo$ in the sense that \cref{eq:sij_simplified} is only non-zero if the following \emph{weak} condition holds: 
\begin{equation}
    E_a < \mtx{E}{max}^{(i)} \equiv \Eg + \frac{m^2}{4 \Eg} \left( \frac{\dSN^2}{(t_i + 2 \dSN) \, t_i} + 1 \right) \, . \label{eq:weak_cond}
\end{equation}
Another possible condition on $E_a$ may follow from demanding that $\denv/\beta < \tdec(t_i^\prime)$.
However, this leads to a complicated inequality of a sixth order polynomial in $E_a$, which we did not attempt to simplify further.

Regarding the remaining number of numerical integrals to be computed, \cref{eq:sij_simplified} is as convenient as expressions found in some previous works but without using any approximations.
In particular, we do not assume highly relativistic ALPs ($\beta \rightarrow 1$) or the asymptotic result for \tdec given in \cref{eq:td_approx}.
When some combination of these assumptions is made, or when the $t \leftrightarrow \tdec$ integration is not performed, we recover the formulae previously derived in the literature~\cite{10.1016/0927-6505(93)90004-W,astro-ph/9601104,Raffelt:1996wa,2109.03244,2205.07896,2205.13549}.

\subsection{Non-instantaneous ALP emission}\label{app:noninst_alp_em}

When finite ALP emission times \tem are considered, the geometry in \cref{fig:geometry} is left unchanged.
As a consequence, \cref{eq:decay_geometry} need not be modified.
However, we have to account for the additional time delay in \cref{eq:time_variable}, which becomes
\begin{equation}
    t \equiv \tdec + \tg + \tem - \dSN {\color{lightgray}/\grc} \, , \label{eq:time_variable_ta}
\end{equation}
which gives rise to the condition $\tdec \geq \tem$, or $\Theta(\tdec - \tem)$ since ALPs cannot decay before they are emitted.

We can then simply replace $t \mapsto t - \tem$ in all equations of \cref{app:inst_alp_em}.
In particular, the ALP decay time now becomes
\begin{align}
    \tdec = \frac{t - \tem + \dSN}{2 \, (1 - \Eg/E_a)} \, \left[1 - \sqrt{1 - \left[1 - \left(\frac{\dSN}{t - \tem + \dSN}\right)^2\right]\frac{4 \Eg (E_a - \Eg)}{\ma^2}}\right] \, . \label{eq:tdec_ta}
\end{align}

Overall, the signal computation becomes slightly more involved, as one more integral (over \tem) appears.
It is convenient to perform this as the innermost integral, keeping in mind that also one new conditions arises in
\begin{equation}
    \mtx{t}{min}^{(i)} = \max\left\{\tem \, , \, \tdec(t_i - \tem) \, , \, \denv/\beta \right\} \; \text{and} \; \mtx{t}{max}^{(i)} = \min\left\{ \tdec(t_i^\prime - \tem) \, , \, \tdec(\mtx{t}{geo}) \right\} \, .
\end{equation}
In the case of SN1987A, $\denv/\beta\grc \geq \denv{\color{gray}/c} > \tem$ such that this new condition is trivial.

\setlength{\bibsep}{0.25ex plus 0.25ex}
\renewcommand{\bibfont}{\footnotesize}
\bibliographystyle{apsrev4-1_mod}
\bibliography{bibliography}

\end{document}